\documentclass[a4paper, 10pt, conference]{ieeeconf}
\usepackage{makecell}
\usepackage{url}

\usepackage{graphicx}

\title{\LARGE \bf
Virtual camera detection: Catching video injection attacks in remote biometric systems
}

\author{\parbox{16cm}{\centering
    {\large Daniyar Kurmankhojayev, Andrei Shadrikov, Dmitrii Gordin, Mikhail Shkorin, Danijar Gabdullin, Aigerim Kambetbayeva, Kanat Kuatov}\\
    {\normalsize
    Department of Research and Development, Verigram, Almaty, Kazakhstan\\}}
}

\begin{document}

\thispagestyle{empty}
\pagestyle{empty}
\maketitle

\begin{abstract}

Face anti-spoofing (FAS) is a vital component of remote biometric authentication systems based on facial recognition, increasingly used across web-based applications. Among emerging threats, video injection attacks—facilitated by technologies such as deepfakes and virtual camera software—pose significant challenges to system integrity. While virtual camera detection (VCD) has shown potential as a countermeasure, existing literature offers limited insight into its practical implementation and evaluation. This study introduces a machine learning-based approach to VCD, with a focus on its design and validation. The model is trained on metadata collected during sessions with authentic users. Empirical results demonstrate its effectiveness in identifying video injection attempts and reducing the risk of malicious users bypassing FAS systems.

\end{abstract}

\section{INTRODUCTION}

\subsection{Background and motivation}

With the growing reliance on web applications for secure access to digital services, robust user authentication mechanisms have become essential. Among various approaches to user authentication, remote biometric systems have gained prominence due to their convenience and adaptability to web-based environments \cite{rui2018}. Among the available biometric modalities, face recognition-based remote biometric systems are particularly favored for their balance of usability and relatively high accuracy \cite{galbally2014}. These systems enable identity verification by analyzing facial features captured through a user's device camera, offering a scalable solution for secure access across digital platforms \cite{abdullahi2024}.

Despite its widespread use, face recognition in remote settings is inherently vulnerable to face spoofing attacks, where adversaries attempt to deceive the system using non-genuine facial inputs \cite{ratha2001, elliott2010, xing2025}. Face spoofing encompasses multiple attack types, including but not limited to presentation attacks (PAs) and video injection attacks. Presentation attacks involve displaying physical artifacts—such as printed photos, replayed videos, or masks—in front of the camera to impersonate a legitimate user \cite{busch2023}. In contrast, video injection attacks bypass the physical camera interface entirely by injecting pre-recorded or synthetic video streams directly into the authentication pipeline, often using deepfake technologies for face swapping and virtual camera software for video injection \cite{carta2022}.

Rapid advancement and public accessibility of face swapping technologies like deepfakes and virtual camera software have exacerbated vulnerabilities in remote biometric authentication systems \cite{antil2025}. Adversaries can easily generate synthetic facial videos—often derived from publicly available images—and inject them through virtual camera software that convincingly mimics real hardware devices. This can impersonate legitimate users and potentially circumvent liveness detection mechanisms, which are designed to verify the presence of a live human subject during authentication \cite{vijaykumar2024}. The threat posed by deepfakes is no longer speculative. Recent reports indicate that 72\% of consumers express daily concerns about being misled by synthetic media \cite{burt2024}, underscoring both the realism and proliferation of such content.

To mitigate these threats, face antispoofing (FAS) methods are integrated as a set of protective layers preceding the recognition process \cite{antil2025}. These methods aim to distinguish between genuine and spoofed facial inputs, thereby enhancing the security of the overall system. Among the various anti-spoofing strategies, liveness detection and virtual camera detection (VCD) serve as two complementary layers within the FAS pipeline \cite{carta2022}. Liveness detection techniques assess physiological or behavioral signals—such as blinking in the eyes, facial movement, or depth information—to determine whether the input originates from a live human subject \cite{khairnar2023}. These methods are effective against many PAs \cite{busch2023}, but may fall short when confronted with sophisticated video injection scenarios, where the spoofed content convincingly mimics live behavior \cite{antil2025}.

\begin{figure}[t]
    \centering
    \includegraphics[scale=0.1]{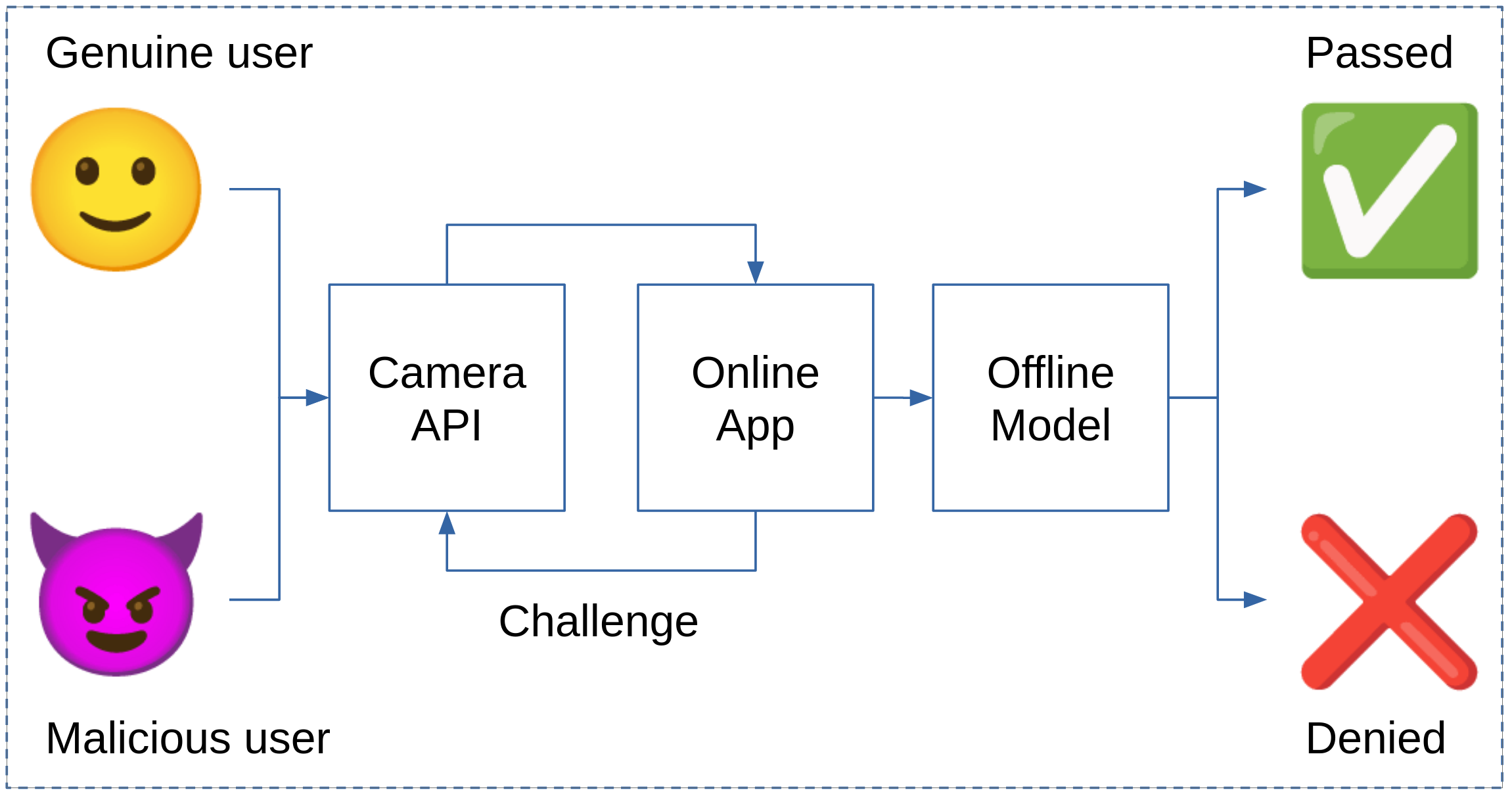}
    \caption{An overview of our setup for detecting malicious users using challenges to their camera API. Our virtual camera detection system does not rely on any visual cue passed to the online application. Rather, we use the browser API to pass several challenges to the camera driver, and the collected results are used to distinguish genuine users from malicious ones.}
    \label{Fig:PipelineOverview}
\end{figure}

VCD, in turn, targets the input source itself by identifying whether the video stream originates from a physical camera or a software-based virtual device. This layer is particularly relevant for countering video injection attacks, which exploit virtual camera tools to feed synthetic or pre-recorded content into the system \cite{carta2022, carta2023}. Despite its effectiveness in addressing such threats, VCD remains an underexplored area in the literature, with limited coverage of practical detection methods tailored to this challenge.

\subsection{Literature review}

The concept of video injection in FAS systems encompasses a range of attack methods, including synthetic video injection via virtual camera software and direct injection downstream of the face recognition pipeline without using a virtual camera \cite{carta2022, carta2023}. According to the biometric system vulnerability model proposed by Ratha et al. (2001) \cite{ratha2001} and expanded by Elliott et al. (2010) \cite{elliott2010}, such attacks can target either the biometric sample acquisition stage or the transmission phase within the pipeline. Carta et al. (2023) \cite{carta2023} identified three layers of protection against these threats: presentation attack detection (PAD), VCD, and IT security safeguards.

This study focuses specifically on detecting video injection attacks carried out using virtual camera software. The existing literature addresses these attacks primarily through two approaches: PAD and VCD, with a predominant emphasis on PAD-based methods. PAD techniques analyze video frames using computer vision, either through manual feature extraction (e.g. texture, facial expressions) \cite{boulkenafet2016, yin2016, nowara2017} or deep learning models \cite{wang2019, chuang2023, yu2024}. For a broader context, comprehensive reviews are available in \cite{ramachandra2017, sharma2023, antil2025}. However, recent advances in face-swapping technologies have significantly improved the realism of deepfakes, making traditional PAD methods increasingly vulnerable and less effective against sophisticated video injection attacks using virtual camera software \cite{burt2024}. Multiple studies show not only a white-box attacks (those in which the malicious user have full knowledge about the technology), but even viable black-box attacks \cite{carlini2020,dong2023}.

In contrast, VCD methods offer a promising alternative to overcome the limitations of PAD-based approaches. Although these methods are already being implemented and widely used in industry to counter video injection attacks, the academic literature on them remains sparse. For example, \cite{carta2023} briefly mention VCD as part of the protection layers in Unissey, a French company's remote biometric system, but do not elaborate on the method itself. To the best of our knowledge, to date, no studies provide a detailed examination of the underlying techniques.

Distinguishing between a physical and virtual camera during a user session is inherently challenging. Virtual cameras are designed to mimic the behavior of physical devices, and any differences are often subtle, context-dependent, and difficult to detect. These distinctions must typically be inferred from metadata collected during authentication, which is constrained by time and system resources. Furthermore, the wide variety of virtual camera software, operating systems, and hardware configurations adds complexity, making it unlikely that a single intuitive feature can reliably differentiate genuine inputs from spoofed ones. This underscores the need for deeper investigation into VCD techniques.

\subsection{Objectives and contributions}
The key objectives of this study can be summarized as follows:
\begin{itemize}
    \item To develop a machine learning-based method for VCD, capable of distinguishing real camera inputs from virtual sources during user authentication sessions.
    \item To identify and extract metadata features that can be feasibly collected during authentication and used to train the detection model.
    \item To empirically validate the proposed method using real-world authentication data, assessing its effectiveness in detecting video injection attempts.
\end{itemize}

The remainder of this paper is organized as follows. Section 2 describes the methodology, including the data collection process, feature extraction, and model development for VCD. Section 3 presents the experimental setup, results, and a discussion of the model’s performance and limitations. Finally, Section 4 concludes the study by summarizing the key objectives, contributions, findings, and insights, and outlines directions for future research.

\section{METHODOLOGY}
This study develops a machine learning-based model for VCD, trained on metadata collected from real-world authentication sessions. The approach is validated through empirical experiments, demonstrating its effectiveness in identifying video injection attacks and contributing to the broader goal of securing remote biometric systems. In this study, we focus on authentication systems implemented within browser-based applications, rather than those native to any specific platform. 

This section begins by outlining the assumptions regarding the types of video injection attacks considered. It then describes the data collection process, followed by an explanation of the proposed detection model.

\subsection{Video injection attacks}

\begin{figure}[thpb]
    \centering
    \includegraphics[scale=0.26]{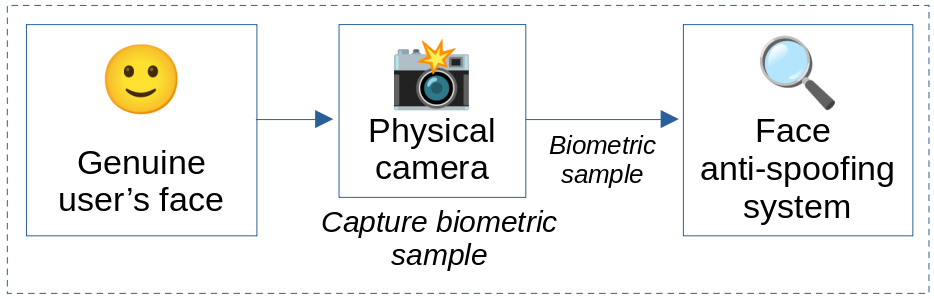}
    \caption{Legitimate biometric session: A genuine user’s face is captured by a physical camera and sent to the FAS system.}
    \label{Fig:RealSession}
\end{figure}

The method developed in this study effectively detects video injection attacks carried out through virtual camera software. To contextualize the approach, we first introduce the concept of an authentication session within a remote biometric system based on face recognition. Figure~\ref{Fig:RealSession} provides a schematic overview of the authentication process in such a system. We then detail the specific types of video injection attacks addressed in this study.

As shown in Figure~\ref{Fig:RealSession}, in a face recognition-based remote biometric system, a user appears in front of the device's camera. This camera collects the biometric sample of the user and sends it to the FAS system, which is usually executed before face recognition itself in order to ensure that a genuine biometric sample is matched against the sample in the database.

\begin{figure}[thpb]
    \centering
    \includegraphics[scale=0.25]{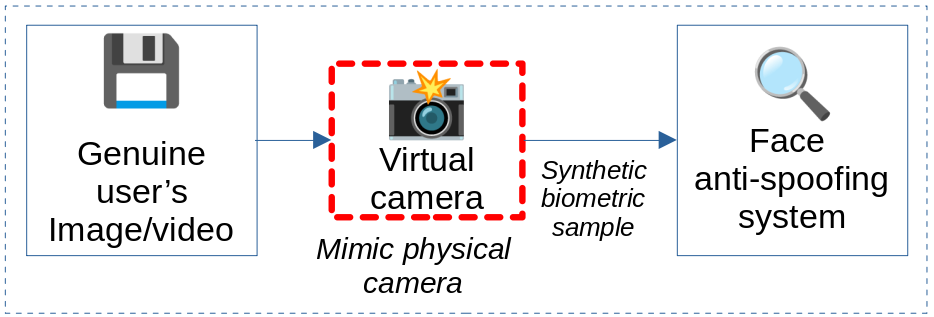}
    \caption{A synthetic biometric sample is sent to the FAS system via a virtual camera used to mimic a physical camera.}
    \label{Fig:Attack1}
\end{figure}

\begin{figure}[thpb]
    \centering
    \includegraphics[scale=0.19]{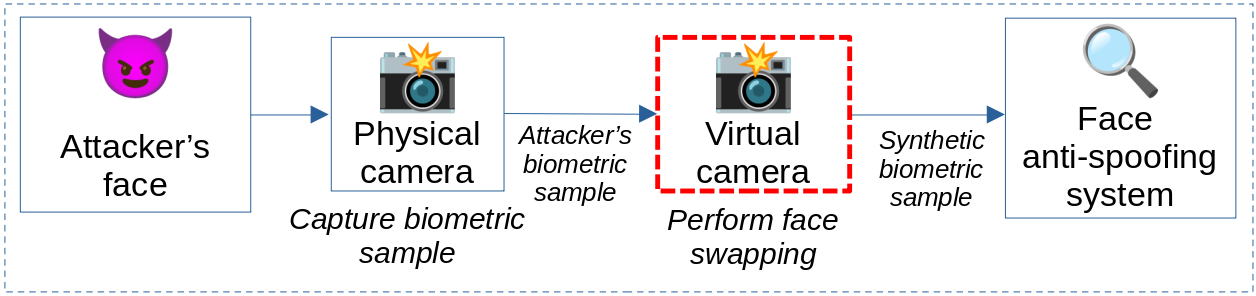}
    \caption{The biometric sample is modified by swapping the attacker's face to the genuine user’s face and sent to the FAS system.}
    \label{Fig:Attack2}
\end{figure}

The following types of video injection attacks can be effectively detected by the method proposed in this study. Figure~\ref{Fig:Attack1} and Figure~\ref{Fig:Attack2} schematically depicts them.

Figure~\ref{Fig:Attack1} shows a simple video injection attack via virtual camera software. In this case, the physical camera is substituted by a virtual camera software, via which an impostor tries to pass the system by showing pre-generated images or videos of the legitimate user. For example, the impostor can get an image of the legitimate user and manipulate it on demand while going through the authentication session (e.g., scale on demand in real-time).

Figure~\ref{Fig:Attack2} shows another type of video injection attack, in which an impostor performs a face swap in real-time using one of the deepfake generating applications. In this case, the data is captured by a real camera and then undergoes additional processing within a virtual camera software.

Below, we describe the data collection process and the models used.

\subsection{Data collection and annotation}

In the client application, we can query the camera’s capabilities and limitations, adjust parameters such as frame rate, frame height, and frame width (among other settings), and measure the response time to these requests. 

\begin{figure}[thpb]
    \centering
    \includegraphics[scale=0.25]{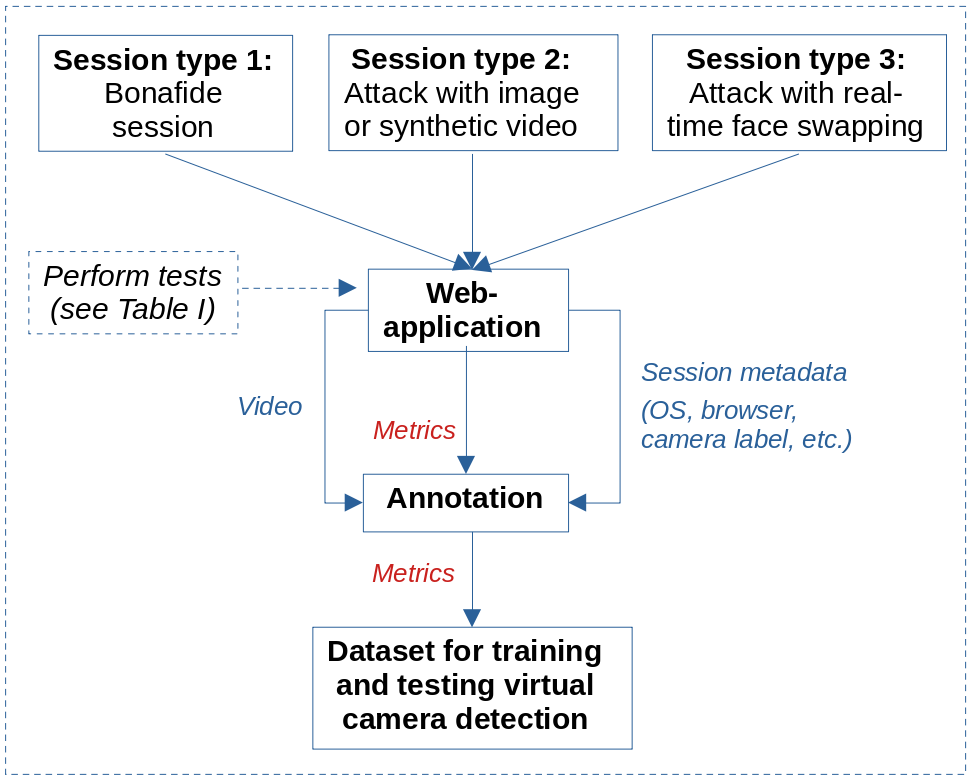}
    \caption{Dataset collection process.}
    \label{Fig:Dataset}
\end{figure}

Figure~\ref{Fig:Dataset} illustrates the overall data collection and annotation process.

As shown in Figure~\ref{Fig:Dataset}, we categorize sessions into three types: legitimate (or bonafide), virtual camera attacks using static images or synthetic videos, and virtual camera attacks involving real-time face swapping. Data collection is conducted via a web application, which executes a series of heuristic-based tests and captures video, session metadata, and performance metrics. The video and metadata are used exclusively for data annotation, while the metrics are specifically employed for training the machine learning model.

The heuristics were developed based on several key observations. Although virtual cameras are designed to emulate physical ones, their behavior under varying configurations can differ markedly. While the response of a virtual camera is influenced by multiple factors—such as the software platform, browser, operating system, and more—certain consistent patterns emerge. For example, reconfiguring a physical camera typically incurs a delay due to its interaction with hardware, whereas a virtual camera processes such changes entirely in software, often more rapidly. In some cases, a virtual camera may indicate that frame settings have been updated, even when no actual changes occur—an implicit modification—whereas physical cameras tend to apply changes explicitly. These behavioral distinctions are among the many factors integrated into our VCD strategy.

Based on these heuristics, we designed a series of tests to generate diagnostic metrics. Table~\ref{Table:Metrics} summarizes the test procedures and the corresponding metrics. During each session, the application performs multiple camera configuration tests to assess behavioral responses. Specifically, we conduct two sets of tests: one targeting frame height and the other targeting frame rate (frames per second). In each case, we collect the requested value, the value reported by the browser, the actual value applied, and the response time. For frame height-related tests, we additionally record the reported and actual frame width, as width may adjust automatically depending on the configuration.

\begin{table}
\caption{Tests and collected metrics (features).}
\label{Table:Metrics}
\begin{center}
\begin{tabular}{ccc}
\hline
\textbf{Set of tests} & \textbf{Requested values} & \textbf{Metrics}\\
\hline
\makecell{Tests on\\frame height} & \makecell{11, 22, 340\\640, 1001, 2001\\3001, 10001} & \makecell{
Requested height\\Reported height\\Actual height\\Reported width\\Actual width\\Response time} \\
\hline
\makecell{Tests on\\frame per second\\(FPS)} &
\makecell{1, 5, 30\\60, 120, 200} &
\makecell{Requested FPS\\Reported FPS\\Actual FPS\\Response time}\\
\hline
\end{tabular}
\end{center}
\end{table}

Cameras may respond differently depending on the settings: standard configurations are often accepted directly, while non-standard ones may be modified. Requested settings can be accepted as-is, explicitly adjusted, or implicitly modified. In explicit adjustments, the browser applies valid settings and reports them directly to the application. In implicit cases, the camera—via the browser—may indicate that changes were applied, while in reality reverting to default values. Because the application operates across diverse platforms, operating systems, browsers, and camera types, responses can vary depending on the browser, operating system, camera driver, and the camera itself. To capture this variability, configurations are systematically varied within each session.

Below, we describe the models used for VCD.

\subsection{Model for virtual camera detection}

VCD is a challenging task that requires machine learning (ML) methods. Due to the wide variety of real cameras, platforms, browsers, devices, and virtual camera software, it is impractical to manually isolate a single factor that reliably distinguishes a virtual camera from a real one. Only a complex combination of features can achieve this with reasonably high accuracy.

\begin{figure}[thpb]
    \centering
    \includegraphics[scale=0.22]{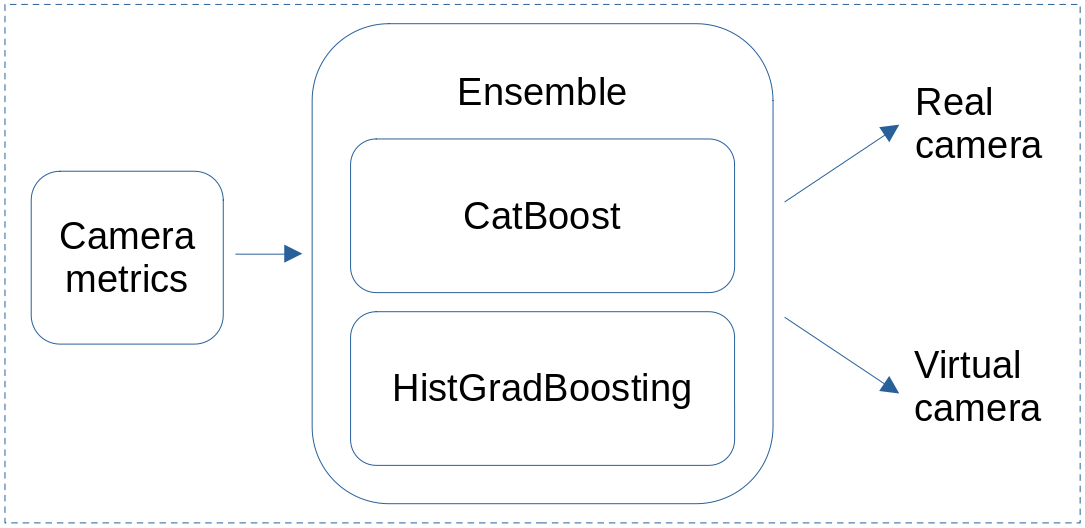}
    \caption{Model for virtual camera detection.}
    \label{Fig:Model}
\end{figure}

Figure~\ref{Fig:Model} provides a visual summary of the model and the training process.

We argue that in our setup, data quality is a higher priority than model complexity. To challenge this claim, we trained multiple models and observed that their performance remained consistently similar. Without loss of generality, we selected three models for this study: Histogram Gradient Boosting (HGB) \cite{guryanov2019}, Categorical Boosting (CatBoost) \cite{prokhorenkova2018}, and an ensemble combining both. These models are well-suited to real-world data challenges such as class imbalance and missing values. CatBoost handles categorical features and missing data natively, requiring no preprocessing, and its ordered boosting technique helps mitigate overfitting—particularly beneficial for imbalanced datasets. HGB employs histogram-based binning to efficiently process large-scale numeric data and assigns missing values to dedicated bins. When combined in an ensemble, these models complement each other’s strengths, enhancing robustness and predictive accuracy across diverse, noisy, and skewed datasets.

\section{NUMERICAL EXPERIMENTS}

In this section, we begin with a descriptive analysis of the dataset to establish its structure and key characteristics. We then present the experimental results derived from our evaluation framework. Finally, we discuss the implications of these findings in the context of FAS and remote biometric authentication based on face recognition.

\subsection{Dataset analysis}
We present the distribution of sessions across platforms, browsers, and virtual camera software observed in the dataset. Table~\ref{Table:Dataset} summarizes platform and browser usage, while Table~\ref{Table:VirCam} details the occurrence of virtual camera tools. This breakdown is essential for assessing the generalizability of our detection methods, identifying prevalent attack vectors, and supporting reproducibility across diverse environments.

\begin{table}
\caption{Distribution of attacks and bonafide sessions among platforms and browsers.}
\label{Table:Dataset}
\begin{center}
\begin{tabular}{ccccc}
\hline
\textbf{Platform} & \textbf{Browser} & \textbf{Attack} & \textbf{Bonafide} & \textbf{Total}\\
\hline
Android	& Chrome	& 923	& 20819	& 21742\\
		& Firefox	& 131	& 301	& 432\\
		& Other	    & 12	& 1949	& 1961\\
\hline
iOS	    & Safari	& 155	& 3815	& 3970\\
		& Chrome	& 0	    & 103	& 103\\
		& Other	    & 0	    & 140	& 140\\
\hline
Linux	& Chrome	& 116	& 3	    & 119\\
		& Firefox	& 94	& 6	    & 100\\
		& Other	    & 15	& 0	    & 15\\
\hline
MacIntel & Chrome	& 125	& 368	& 493\\
		& Safari	& 32	& 124	& 156\\
		& Firefox	& 23	& 6	    & 29\\
		& Other	    & 1	    & 125	& 126\\
\hline
Win32	& Chrome	& 883	& 1994	& 2877\\
		& Firefox	& 264	& 95	& 359\\
		& Other	    & 38	& 152	& 190\\
\hline
		& \textbf{Total}	    & \textbf{2812}	& \textbf{30000}	& \textbf{32812}\\
\hline  
\end{tabular}
\end{center}
\end{table}

As shown in Table~\ref{Table:Dataset}, the dataset comprises over 30,000 sessions collected across various platforms—including Android, iOS, Linux, MacIntel, and Win32—and browsers, predominantly Chrome and Firefox. It includes both bonafide and attack sessions. Bonafide sessions were most frequently recorded from Android devices, followed by iOS and Win32. Attack sessions accounted for approximately 10\% of the bonafide volume. Notably, the majority of attacks originated from the Win32 platform, followed by Android, Linux, MacIntel, and iOS. In terms of platform-browser combinations, the highest number of attacks was observed from Android-Chrome, followed by Win32-Chrome and Win32-Firefox.

\begin{table}
\caption{Virtual camera software identified in the dataset.}
\label{Table:VirCam}
\begin{center}
\begin{tabular}{cc}
\hline
\textbf{Virtual camera software} & \textbf{Count}\\
\hline
OBS Studio & 251 \\
SplitCam & 215 \\
ManyCam & 190 \\
Iriun Webcam & 54 \\
Other & 24 \\
Unknown & 2078 \\
\hline  
\textbf{Total} & \textbf{2812}\\
\hline  
\end{tabular}
\end{center}
\end{table}

As part of the metadata, we collected camera labels, which in some cases revealed the virtual camera software used during attack sessions. However, for the majority of sessions, this information was either omitted or deliberately obfuscated—often replaced with custom names or descriptors indicating camera orientation (e.g., “front” or “back”). This limits the reliability of camera labels as a standalone indicator for VCD. For this reason, we did not use it as a feature. Despite these inconsistencies, several hundred sessions did include identifiable software names. As shown in Table~\ref{Table:VirCam}, the most frequently observed virtual camera tools were OBS Studio, SplitCam, ManyCam, and Iriun Webcam.

Next, we examine a specific scenario involving the Linux platform, Chrome browser, and OBS Studio as the virtual camera software. Table~\ref{Table:RealSession} and Table~\ref{Table:AttackSession} present representative bonafide and attack sessions under this configuration. In this case, the heuristics outlined in the previous section successfully distinguished between real and virtual camera behavior, demonstrating their effectiveness in practical settings.

As shown in Table~\ref{Table:RealSession}, the frame height changed as requested throughout the tests, up to the final value of 3001, at which point the resolution was set to 1920×1080. Notably, the application times were heterogeneous, suggesting that standard frame settings were applied more quickly than non-standard ones. In contrast, Table~\ref{Table:AttackSession} shows that the frame height changed as requested only up to the fourth test (i.e., when the requested height was 640), after which no further changes occurred. This likely reflects a maximum resolution constraint configured in OBS Studio for that session, although higher settings are technically possible. Such constraints illustrate how configuration limits can manifest in virtual camera behavior and may aid in detection. Interestingly, the application times in this attack session were relatively low and consistent.

\begin{table}
\caption{A session with a real camera (Linux-Chrome).}
\label{Table:RealSession}
\begin{center}
\begin{tabular}{ccc}
\hline
\textbf{Requested height} & \textbf{W x H} & \textbf{Apply time}\\
\hline
11	    & 20x11	    & 210\\
22	    & 39x22	    & 1.9\\
240	    & 320x240	& 94.5\\
640	    & 1138x640	& 91.7\\
1001	& 1780x1001	& 92.2\\
3001	& 1920x1080	& 1.7\\
\hline
\end{tabular}
\end{center}
\end{table}

\begin{table}
\caption{A session with OBS camera (Linux-Chrome).}
\label{Table:AttackSession}
\begin{center}
\begin{tabular}{ccc}
\hline
\textbf{Requested height} & \textbf{W x H} & \textbf{Apply time}\\
\hline
11	    & 20x11	    & 0.8\\
22	    & 39x22	    & 1.3\\
240	    & 427x240	& 1.6\\
640	    & 640x360	& 1.6\\
1001	& 640x360	& 1.2\\
3001	& 640x360	& 1.3\\
\hline
\end{tabular}
\end{center}
\end{table}

\subsection{Results}

In VCD, the goal is to identify whether a video feed is coming from a real physical camera or a virtual/synthetic source (like OBS, ManyCam, or software-generated streams). To evaluate the effectiveness of the proposed VCD approach, we trained and tested three models: CatBoost, HGB, and an Ensemble model combining both. The dataset was partitioned into training (60\%), validation (20\%), and testing (20\%) subsets, ensuring a balanced distribution of bonafide and attack sessions across all splits.

Model performance was assessed using standard face anti-spoofing metrics:
\begin{itemize}
    \item \textit{AUC-ROC} (Area Under the Receiver Operating Characteristic Curve)
    \item \textit{APCER} (Attack Presentation Classification Error Rate)
    \item \textit{BPCER} (Bona Fide Presentation Classification Error Rate)
    \item \textit{ACER} (Average Classification Error Rate)
\end{itemize}

These metrics provide a comprehensive view of each model’s ability to distinguish between legitimate and spoofed sessions.

\begin{figure}[t]
    \centering
    \includegraphics[scale=0.3]{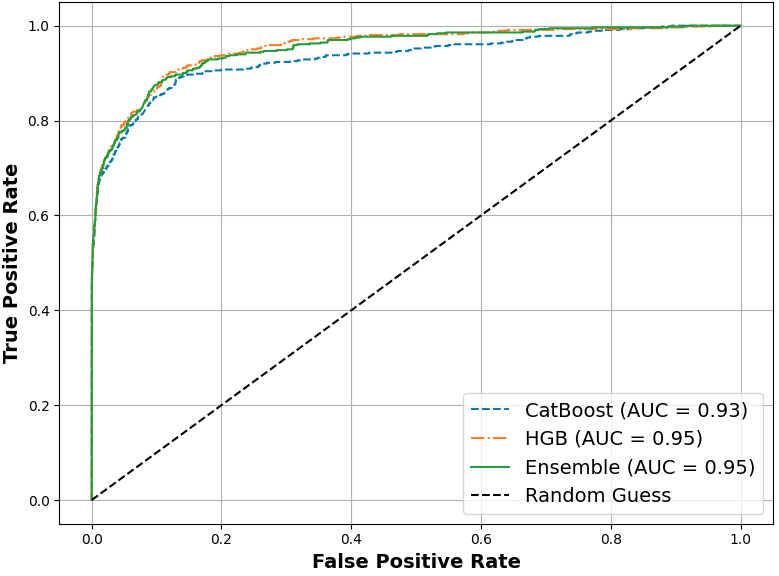}
    \caption{AUC-ROC curves for CatBoost, HGB, and Ensemble models}
    \label{Fig:ModelROC}
\end{figure}

Figure \ref{Fig:ModelROC} shows AUC-ROC curves. All three models — CatBoost, HGB, and the Ensemble — demonstrated strong discriminative capability, each achieving an AUC-ROC above 0.9. This high area under the ROC curve indicates that the models are highly effective at ranking bonafide (real camera) inputs above virtual (synthetic) ones across a wide range of thresholds. In other words, the classifiers have learned meaningful patterns that separate real and virtual camera feeds with high confidence, regardless of the specific decision boundary.

\begin{figure}[t]
    \centering
    \includegraphics[scale=0.3]{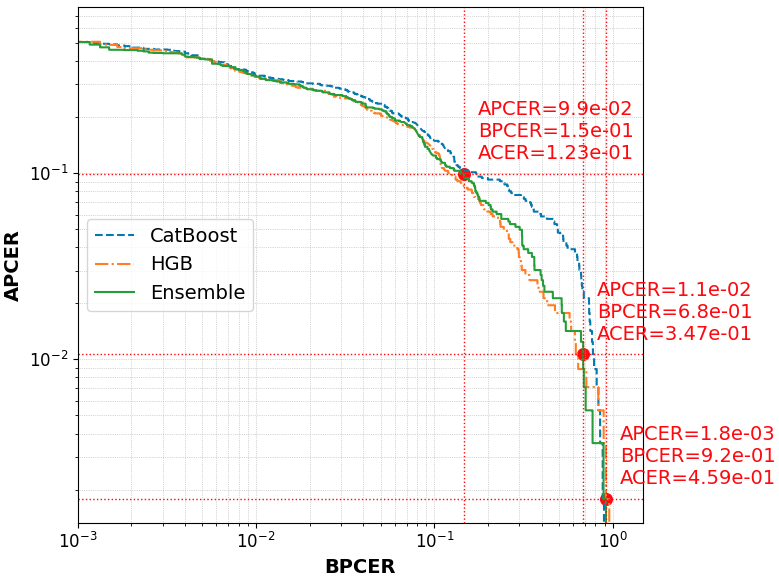}
    \caption{DET curves for CatBoost, HGB, and Ensemble models.}
    \label{Fig:ModelDet}
\end{figure}

However, while AUC-ROC reflects the model’s overall potential, it does not account for the practical trade-offs that arise when a fixed threshold is applied. To evaluate real-world usability, we analyzed system performance at three fixed APCER levels: $10^{-1}$, $10^{-2}$, and $10^{-3}$, corresponding to 10\%, 1\%, and 0.1\% of virtual camera inputs being falsely accepted as bonafide. Specifically, the detection error tradeoff (DET) curves were generated to visualize sensitivity across operating thresholds (Figure \ref{Fig:ModelDet}).

As illustrated in Figure \ref{Fig:ModelDet}, we selected three target APCER levels: $10^{-1}$, $10^{-2}$, and $10^{-3}$. These correspond to 10\%, 1\%, and 0.1\% of virtual camera inputs being incorrectly classified as bonafide, respectively.

Table \ref{Table:apcer_tradeoff} presents the system’s performance at these three fixed APCER levels. These thresholds reflect increasing levels of security enforcement, with progressively stricter rejection of spoofed or synthetic camera feeds.

As the APCER decreases, the system becomes more effective at identifying and blocking virtual camera sources. However, this improvement in security comes at the cost of usability, as evidenced by the rising BPCER values — the rate at which genuine camera inputs are mistakenly rejected.

At $10^{-1}$ APCER, the system maintains a relatively low BPCER of 14.6\%, indicating that most real users are correctly accepted. This configuration offers a balanced trade-off, suitable for environments where moderate security is sufficient and user experience is a priority.

When the APCER is tightened to $10^{-2}$, the BPCER escalates sharply to 68.3\%. This means that more than two-thirds of genuine users are falsely rejected. Such a high rejection rate introduces noticeable usability degradation, especially in real-world applications like remote authentication, video conferencing, or online proctoring. Users may be repeatedly denied access despite using legitimate camera hardware, leading to frustration, increased support burden, and potential abandonment of the system.

At the strictest level, $10^{-3}$, the BPCER reaches 91.7\%, indicating that nearly all genuine users are misclassified as virtual. While this configuration offers maximum protection against spoofing, it renders the system practically unusable for legitimate users. The severe usability loss at this level makes it suitable only for highly sensitive environments where security requirements outweigh operational accessibility --- for example, in forensic analysis or high-assurance identity verification.

\begin{table}[t]
\centering
\caption{System performance at fixed APCER levels.}
\label{Table:apcer_tradeoff}
\begin{tabular}{cccc}
\hline
\textbf{APCER} & \textbf{BPCER} & \textbf{ACER} & \textbf{Interpretation} \\
\hline
$10^{-1}$ & 14.6\% & 12.3\% & \makecell{Balanced configuration\\with moderate security\\and good usability} \\
\hline
$10^{-2}$ & 68.3\% & 34.7\% & \makecell{Stronger security with\\noticeable usability degradation} \\
\hline
$10^{-3}$ & 91.7\% & 45.9\% & \makecell{Maximum security with\\severe usability loss} \\
\hline
\end{tabular}
\end{table}

It is worth noting that training on a larger dataset than employed in this study, with identical model configurations, resulted in noticeably improved performance.

\subsection{Discussion}

The results indicate that the trained model performs effectively in detecting virtual camera usage, achieving high accuracy across a broad and diverse dataset. The dataset, which includes over 30,000 sessions spanning multiple platforms, browsers, and camera configurations, reflects realistic deployment conditions in remote biometric authentication systems. This diversity contributes to the model’s robustness and its ability to generalize across heterogeneous environments.

The model’s performance against virtual camera attacks is particularly noteworthy. Despite the variability in camera metadata and the presence of obfuscated or custom labels, the model consistently identified virtual camera behavior. This suggests that the features and heuristics used during training — such as resolution constraints, response time patterns, and configuration anomalies — are sufficiently discriminative to support reliable detection.

Moreover, the model maintained a low false positive rate in bonafide sessions, indicating that it can be deployed as a pre-screening layer without significantly impacting user experience. This is especially valuable in remote authentication scenarios, where user interaction is limited and real-time decisions are required.

To further assess operational viability, we examined the trade-off between APCER (false acceptance of virtual cameras) and BPCER (false rejection of real cameras) at fixed thresholds. As expected, tightening the APCER from $10^{-1}$ to $10^{-3}$ significantly reduced the likelihood of virtual camera spoofing but sharply increased BPCER — reaching up to 91.7\% at the strictest setting. This reflects a classic security–usability tension: while lower APCER enhances protection, it risks excluding legitimate users, which can be unacceptable in user-facing systems.

However, it is important to contextualize this trade-off within a multi-layered face anti-spoofing framework. VCD is typically one of several protective layers, alongside liveness detection, challenge–response mechanisms, and behavioral analysis. In such architectures, it may be viable — and even preferable — to operate VCD at a higher APCER, accepting a small number of missed attacks at this stage. These can then be intercepted by downstream modules that specialize in detecting replay attacks, deepfakes, or synthetic content. This layered approach allows for more flexible thresholding, balancing security and usability across the entire pipeline rather than within a single module.

Overall, these findings support the integration of VCD as a meaningful component of anti-spoofing systems. While not sufficient on its own, it strengthens the overall defense posture and improves resilience against increasingly sophisticated attack vectors. Future work may explore adaptive thresholding strategies, ensemble decision-making across layers, and mobile-specific evaluation, where hardware constraints and attack surfaces differ substantially.

\section{CONCLUSIONS}

This study investigated the use of virtual camera detection (VCD) as a protective layer in remote face recognition-based authentication systems. A machine learning model was developed and trained on a broad dataset comprising over 30,000 sessions across diverse platforms, browsers, and camera configurations. The model achieved high accuracy in identifying virtual camera usage, demonstrating its effectiveness in mitigating video injection attacks under realistic conditions.

The scope of this work was limited to attacks involving virtual camera software. Other forms of video injection—such as feed overwriting or phishing-based session hijacking—were not addressed, as they bypass the virtual camera interface and are more effectively mitigated through software-level restrictions and secure infrastructure. While it was theoretically possible for attackers to manipulate metadata or overwrite request results, doing so required domain expertise and effort that likely exceeded the practical benefit. Additionally, VCD was evaluated as a standalone protection layer; in practice, it should be integrated with liveness detection and broader IT security measures to ensure comprehensive defense.

Future work should primarily focus on advancing VCD itself. One direction is to further improve detection methods by incorporating richer metadata features, exploring temporal patterns, and applying adaptive learning techniques to counter evolving virtual camera software. Enhancing robustness against deliberate obfuscation and metadata manipulation will be critical for maintaining detection reliability in adversarial settings. The second direction involves integrating VCD with complementary protection layers, such as liveness detection and secure session protocols, to address broader attack scenarios—including those that bypass virtual camera interfaces. This layered approach will help mitigate the limitations identified in this study and strengthen the overall resilience of remote biometric authentication systems.

\section{ACKNOWLEDGMENTS}

During the preparation of this work, the author(s) used the AI language model Copilot, developed by Microsoft, to assist with grammar and clarity. All content was subsequently reviewed and edited by the author(s), who take full responsibility for the final version of the publication.
 
{\small
\bibliographystyle{ieeetr}      
\bibliography{egbib}
}

\end{document}